\documentclass[journal]{IEEEtran}
\usepackage{tensor}
\usepackage{graphicx}
\usepackage{subcaption}
\usepackage{tabularx}
\usepackage[cmex10]{amsmath}
\usepackage{amsthm}
\usepackage{amssymb}
\usepackage{bm}
\usepackage{algorithm}
\usepackage{algorithmic}
\usepackage{cite}
\usepackage{setspace}
\usepackage{stfloats}
\usepackage{enumerate}
\usepackage{cases}
\usepackage{multirow}
\usepackage{mathrsfs}
\usepackage{array}
\usepackage{color}
\usepackage{verbatim}
\usepackage{extpfeil} 
\usepackage{booktabs}
\usepackage{multirow}
\usepackage{caption}
\usepackage{graphicx}
\usepackage{tabularx}
\usepackage{epstopdf}
\usepackage{textcomp}
\usepackage{epsfig,epsf,color,balance,cite}
\usepackage{amssymb}
\usepackage{amsthm}
\usepackage{graphicx}
\usepackage{array,color}
\usepackage{amsmath}
\usepackage{graphicx}
\usepackage{tabularx}
\usepackage{epsfig,epsf,color,balance,cite}
\usepackage{algorithmic}
\usepackage{algorithm}
\usepackage{bm}
\usepackage{caption}
\usepackage{textcomp}
\usepackage{color}
\usepackage{multirow}
\usepackage{cite}
\usepackage{enumerate}
\usepackage{cases}
\usepackage{color}
\usepackage{epstopdf}
\usepackage{textcomp}
\usepackage{subcaption}
\usepackage{url}
\usepackage{tikz}  
\usepackage[colorlinks]{hyperref}

\def\BibTeX{{\rm B\kern-.05em{\sc i\kern-.025em b}\kern-.08em
		T\kern-.1667em\lower.7ex\hbox{E}\kern-.125emX}}
\begin{document}
	
	\makeatletter
	\newcommand{\rmnum}[1]{\romannumeral #1}
	\newcommand{\Rmnum}[1]{\expandafter \@slowromancap \romannumeral #1@}
	\makeatother
	
	\title{Exploring Dual-Sniffer Passive Localization: Algorithm Design and Experimental Results}
	
	\author{Tuo Wu, Lingyu Hou, Hong Niu, Saihua Xu,\\ Sirajudeen Gulam Razul, Chau Yuen, \emph{Fellow, IEEE}
		\thanks{\emph{(Corresponding author: Chau Yuen.)}}
		
		\thanks{T. Wu, L. Hou, H. Niu, S. Xu and C. Yuen are with the School of Electrical and Electronic Engineering, Nanyang Technological University, 639798, Singapore (E-mail: $\rm \{tuo.wu, e230033, hong.niu, SHXU, chau.yuen\}@ntu.edu.sg$).}
		\thanks{S. G. Razul is with the Temasek Laboratories, Nanyang Technological University, Singapore 637553 (E-mail: $\rm  esirajudeen@ntu.edu.sg$).}
	}
	\markboth{}
	{Lai \MakeLowercase{\textit{et al.}}: }
	
	\maketitle
	
	\thispagestyle{empty}
	
	\begin{abstract}
		In this paper, we explore a dual-sniffer passive localization system that detects the timing difference of signals from both commercial base station  (eNb) and user equipment (UE) to the sniffers. We design two localization schemes for UE localization: a time of arrival (ToA) based scheme and a time difference of arrival (TDoA) based scheme. In the ToA-based scheme, we derive two ellipse equations from measured arrival times at two sniffers, enabling direct numerical computation of the estimated position. For the TDoA-based scheme, we relocate one sniffer to a different position to obtain two sets of TDoA measurements, resulting in hyperbola equations. We then apply a least squares (LS) algorithm to analytically estimate the UE's position. Simulation results validate the effectiveness of the proposed TDoA-based scheme, demonstrating improved accuracy in UE positioning.We build a platform based on the considered localization system and conduct real-world experiments. The experimental results confirm the accuracy and practicality of the TDoA-based dual-sniffer localization scheme, demonstrating improved precision in passive localization.
		
	\end{abstract}
	
	\section{Introduction}
	In the era of fifth-generation (5G) technology, there is an escalating need to accurately locate mobile devices such as smartphones within densely populated urban environments. Applications like navigation \cite{Lu1}, security \cite{Niu1},  and resource allocation \cite{YC1} all rely on precise user equipment (UE) localization \cite{tian2019passive}.
	
	Software-defined radio (SDR) sniffers have emerged as vital tools for UE localization in both long-term evolution (LTE) and 5G networks. software-defined radio (SDR) sniffers have emerged as a vital tool for UE localization. Despite the rollout of 5G, LTE networks continue to be widely deployed and integrated within the existing infrastructure, making LTE sniffers still highly relevant for localization tasks in contemporary networks. However, existing open-source LTE sniffers, such as LTEye \cite{karaccay2021network}   and FALCON \cite{falkenberg2019falcon}, primarily focus on decoding the downlink control channel and are incapable of capturing IP packets or critical cellular protocol packets like radio resource control (RRC) and non-access stratum (NAS). This limitation significantly reduces their applicability in scenarios that require full packet access, particularly in internet of things (IoT) networks where comprehensive monitoring and analysis are crucial \cite{Wu1}.
	
	Fortunately, Kotuliak et al. in \cite{kotuliak2022ltrack} introduced LTE Sniffer, an open-source tool suitable for UE localization in conjunction with commercial base stations (eNbs). Their approach involves mobile fingerprinting by extracting the international mobile subscriber identity (IMSI) of the UE using a dedicated database and a machine learning-based \cite{Wu2} IMSI extractor. However, this method necessitates building a database and consumes substantial resources. In applications such as security, where immediate localization of mobile devices via eNbs is essential, relying on extensive databases and resource-intensive processes poses a non-trivial challenge.
	
	To address this challenge, we introduce a dual-sniffer system that detects the timing difference of signals from both the commercial eNb and the UE to the sniffers. By formulating a localization problem based on these time differences, we can accurately detect and locate the UE without the need for extensive databases or resource-intensive computations.  The main contributions can be summarized as follows:
	\begin{itemize}
		\item We investigate a dual-sniffer localization system that detects the timing difference of signals from both the commercial eNb and the UE to the sniffers.
		\item We design two localization schemes for UE positioning: a time of arrival (ToA) based scheme and a time difference of arrival (TDoA) based scheme. In the ToA-based scheme, we derive two ellipse equations from the measured arrival times at the two sniffers, enabling direct numerical computation of the estimated position.
		\item For the TDoA based scheme, we relocate one sniffer to a different position to obtain two sets of TDoA measurements, resulting in two hyperbola equations. We then apply a least square (LS) algorithm to analytically estimate the position.
		\item We build the platform based on the considered localization system and conduct real experiments, which  validates  the effectiveness of the proposed TDoA based dual-sniffer localization scheme, demonstrating improved accuracy in passive localization.
	\end{itemize}

	\section{System Model}
	In a dual-sniffer localization system, the primary components include the UE, an  eNb, and 2 sniffers (SNs). The eNb, serving as the main connection point for the UE within the LTE network, is located at a fixed and known position denoted by \( \mathbf{p} = [x_p, y_p]^\mathrm{T} \). The true location of the UE is represented by \( \mathbf{u} = [x_u, y_u]^\mathrm{T} \), which is unknown and requires estimation.
	
	It is assumed that the distance between the UE and the eNb is \( d_{\text{UB}} \). Each sniffer, indexed by \( k \) where \( k \in \{1, 2\} \), is strategically placed at a location \( \mathbf{s}_k = [x_k, y_k]^\mathrm{T} \). The distances from the \( k \)-th sniffer to the eNb and to the UE are denoted by \( d_{\text{eNb},k} \) and \( d_{\text{UE},k} \), respectively.  These distances are calculated using the Euclidean distance formula as 
	\begin{align} 
		d_{\text{eNb},k} &= \|\mathbf{s}_k - \mathbf{p}\|= \sqrt{(x_k - x_p)^2 + (y_k - y_p)^2}, \label{d1} \\
		d_{\text{UE},k} &= \|\mathbf{s}_k - \mathbf{u}\| = \sqrt{(x_k - x_u)^2 + (y_k - y_u)^2}, \label{d2} \\
		d_\text{UB} &= \|\mathbf{u} - \mathbf{p}\|= \sqrt{(x_u - x_p)^2 + (y_u - y_p)^2}. \label{d3}
	\end{align} 
	The sniffers are equipped to capture LTE signals transmitted between the eNb and the UE. The UE's location is estimated based on the propagation time of these signals and the TA command \footnote{The TA command, issued by the eNb, compensates for the signal propagation delay between the UE and the eNb by adjusting the timing of the uplink signals from the UE, ensuring synchronization with the eNb's receiving window.}.

	\subsection{Transmission from eNb to UE}
	
	In the downlink transmission from the eNb to the UE, the eNb serves as the reference clock for synchronization. Specifically, the eNb transmits radio frames every $10$\,ms and subframes every every $1$\,ms with negligible drift \footnote{The eNb employs a free-running clock with a drift rate of approximately $2.5$ parts per billion (ppb). For comparison, an oven-controlled crystal oscillator (OCXO) has a drift rate of about $25$ ppb, making the drift of the eNb effectively negligible.}. Although the UE may have a lower-quality or less accurate internal clock, it synchronizes to the eNb's clock to ensure accurate reception of the time-sensitive downlink signals.
	
	Let the transmission times from the eNb be defined as $T_{\text{BS}}$,  representing the subframe timings, which can be formulated as
	\begin{equation}\label{1}
		T_{\text{BS}} = \{t_0, t_1, \cdots, t_n\},
	\end{equation}
	where $n \, (n \in \{1, 2, \cdots, N\})$ is an integer representing the subframe index, and $N$ denotes the total number of transmitted subframes.
	
	Assuming the UE has already received the TA command from the eNb, it transmits uplink frames earlier than it receives the downlink frames to compensate for propagation delays. Specifically, if the eNb transmits a downlink frame at time $T_{\text{BS}} = t_n$, the UE transmits the corresponding uplink frame at 
	\begin{equation}\label{2}
		T_{\text{UE}} = t_n + \frac{d_\text{UB}}{c} - \delta_\text{TA} + \varepsilon_{\text{UE}},
	\end{equation}
	where $c$ denotes the speed of light, $\delta_\text{TA}$ is the timing advance value provided by the eNb, and $\varepsilon_{\text{UE}}$ represents any hardware timing error in the UE. It is important to note that $\delta_{\text{UE}}$ is not a timing offset because it is independent of the eNb's clock.
	
	\subsection{Transmission from eNb/UE to SNs}
	
	Next, we model the transmissions between the eNb and SNs, and between the UE and SNs. Notably, the SNs are not required to be synchronized with GPS for practical localization purposes.
	
	For the eNb-to-SN link, the $k$-th SN receives the downlink frame with respect to (w.r.t.) its own clock at $T_{\text{DL},k}$, which is formulated as
	\begin{equation}\label{3}
		T_{\text{DL},k} = t_n + \frac{d_{\text{eNb},k}}{c}+\delta_{\text{SN},k},
	\end{equation}
	where $\delta_{\text{SN},k}$ represents the time offset of the $k$-th SN w.r.t. the eNb clock.
	
	For the UE-to-SN link, the $k$-th SN receives the uplink frame w.r.t. its own clock at $T_{\text{UL},k}$, which is expressed as 
	\begin{equation}\label{4}
		T_{\text{UL},k}= T_{\text{UE}}+ \frac{d_{\text{UE},k}}{c}+\delta_{\text{SN},k},
	\end{equation}
	Substituting $T_{\text{UE}}$ from  \eqref{2} into  \eqref{4}, we obtain 
	\begin{align}\label{5}
		T_{\text{UL},k}= t_n + \frac{d_\text{UB}}{c} - \delta_\text{TA} + \varepsilon_{\text{UE}}+ \frac{d_{\text{UE},k}}{c}+\delta_{\text{SN},k}.
	\end{align} 
	Since the UE's clock is synchronized to the eNb's clock, the reference to the UE's clock can be neglected in timing calculations.
	
	\subsection{Downlink $\&$ Uplink Subframe Timing Difference}
	Then, let us further model the  downlink $\&$ uplink subframe timing difference, denoted as $\Delta_{k}$, which can be measured  by the open source code. Specifically, we have
	\begin{align}\label{6}
		\Delta_{k}&= T_{\text{DL},k}-T_{\text{UL},k}+\varepsilon_{\text{SN},k}\nonumber\\
		&=\frac{d_{\text{eNb},k}}{c}-\frac{d_\text{UB}}{c}-\frac{d_{\text{UE},k}}{c}+\delta_\text{TA}-\varepsilon_{\text{UE}}+\varepsilon_{\text{SN},k},
	\end{align}
	where $\varepsilon_{\text{SN},k}$ denotes the measurement error at the $k$-th SN. 
	
	Reformulating \eqref{6}, we have the following equation:
	\begin{align}\label{7}
		\frac{d_\text{UB}}{c}+\frac{d_{\text{UE},k}}{c} =\frac{d_{\text{eNb},k}}{c} +\Delta_{k}+\delta_\text{TA}-\varepsilon_{\text{UE}}+\varepsilon_{\text{SN},k},
	\end{align}
	This equation establishes a relationship between the distances involved and can be utilized to derive the estimated position of the UE.
	
	\section{Localization Scheme Design}   
	In this section, we design two kinds of localization scheme for estimating the unknown position of the UE.  
	\subsection{ToA Based Localization Scheme}  
	In this subsection, we present a ToA based localization scheme to estimate the unknown position of the UE. We utilize the timing difference measurements $\Delta_{k}$ from two SNs while ignoring hardware timing errors and measurement noise for simplicity \cite{Smith}.
	
	Starting from \eqref{7}, and neglecting the error terms \(\varepsilon_{\text{UE}}\) and \(\varepsilon_{\text{SN},k}\), we obtain 
	\begin{equation}\label{8}
		\frac{d_{\text{UB}}}{c} + \frac{d_{\text{UE},k}}{c} \approx \frac{d_{\text{eNb},k}}{c} + \Delta_{k} + \delta_\text{TA},
	\end{equation}
	where \( k = 1, 2 \) represents the index of the sniffers. 
	Then, by multiplying both sides of \eqref{8} by \( c \) yields 
	\begin{equation}\label{9}
		d_{\text{UB}} + d_{\text{UE},k} \approx d_{\text{eNb},k} + c (\Delta_{k} + \delta_\text{TA}).
	\end{equation}
	Let us further define $D_{k} = d_{\text{eNb},k} + c (\Delta_{k} + \delta_\text{TA})$, we have  
	\begin{equation}\label{11}
		d_{\text{UB}} + d_{\text{UE},k} \approx D_{k}.
	\end{equation}
	Substituting the expressions for \( d_{\text{UB}} \) and \( d_{\text{UE},k} \) in terms of the positions, we have
	\begin{equation}\label{12}
		\| \mathbf{u} - \mathbf{p} \| + \| \mathbf{u} - \mathbf{s}_k \| \approx D_{k},
	\end{equation} 
	which represents that the sum of distances from the UE to the eNb and to the \( k \)-th sniffer equals a known constant \( D_{k} \). This forms the basis of the ellipse intersection problem. Let us further assume two sniffers (\( k = 1, 2 \)), we obtain two equations, which is given as  \cite{Smith}
	\begin{align}
		\| \mathbf{u} - \mathbf{p} \| + \| \mathbf{u} - \mathbf{s}_1 \| &\approx D_{1}, \label{15} \\
		\| \mathbf{u} - \mathbf{p} \| + \| \mathbf{u} - \mathbf{s}_2 \| &\approx D_{2}. \label{16}
	\end{align} 
	These two ellipse equations can be used to estimate the UE's position \( \mathbf{u} = [x_u, y_u]^\mathrm{T} \) with direct numerical computation.
	
	\subsection{TDoA Based Localization Scheme}    
	In the previous subsection, we directly ignored hardware timing errors and measurement noise to simplify the localization problem. However, in practical measurements, these errors cannot be neglected, as they significantly affect localization accuracy. To further reduce the impact of these errors, we consider using a  TDoA based localization scheme.
	
	Specifically, we first perform measurements with two SNs simultaneously. Then, keeping one SN fixed, we move the other to a different location. This approach allows us to collect measurements from different SN configurations while maintaining synchronization within the same time slot.
	
	From \eqref{7}, including the error terms, and considering the first measurement configuration with SNs at positions \( \mathbf{s}_1 \) and \( \mathbf{s}_2 \), we have
	\begin{align}\label{17}
		\frac{d_{\text{UB}}}{c} + \frac{d_{\text{UE},1}}{c} &= \frac{d_{\text{eNb},1}}{c} + \Delta_{1} + \delta_{\text{TA}} - \varepsilon_{\text{UE}} + \varepsilon_{\text{SN},1}, \\
		\frac{d_{\text{UB}}}{c} + \frac{d_{\text{UE},2}}{c} &= \frac{d_{\text{eNb},2}}{c} + \Delta_{2} + \delta_{\text{TA}} - \varepsilon_{\text{UE}} + \varepsilon_{\text{SN},2}. \label{18}
	\end{align} 
	By subtracting   \eqref{18} from   \eqref{17}, we eliminate common terms such as \( d_{\text{UB}}/c \), \( \delta_{\text{TA}} \), and \( \varepsilon_{\text{UE}} \), resulting in a TDoA expression as 
	\begin{align}\label{19}
		\Delta d_{12}&= {d_{\text{UE},1} - d_{\text{UE},2}} \nonumber\\
		&= {d_{\text{eNb},1} - d_{\text{eNb},2}}  + c(\Delta_{1} - \Delta_{2}) + c(\varepsilon_{\text{SN},1} - \varepsilon_{\text{SN},2})\nonumber\\
		&= d_{\text{eNb},12}   +\delta_{1q}+\varepsilon_{1q},
	\end{align}
	where $d_{\text{eNb},12}=d_{\text{eNb},1} - d_{\text{eNb},2}$, $\delta_{12}= c (\Delta_{1}^\prime - \Delta_{2})$, and $\varepsilon_{12}=c(\varepsilon_{\text{SN},1} - \varepsilon_{\text{SN},2})$.
	
	Next, we move the $2$-nd SN to a new location, at position \( \mathbf{s}_3 \), while keeping the $1$-st SN at \( \mathbf{s}_1 \). We perform another set of measurements, and similar to \eqref{19}, we obtain another TDoA expression as 
	\begin{align}\label{22}
		\Delta d_{12}&= d_{\text{UE},1} - d_{\text{UE},3}\nonumber\\
		&=  d_{\text{eNb},1} - d_{\text{eNb},3} + c(\Delta_{1}^\prime - \Delta_{3})  + c(\varepsilon_{\text{SN},1} - \varepsilon_{\text{SN},3})\nonumber\\
		&=d_{\text{eNb},13} +\delta_{13}+\varepsilon_{13},
	\end{align}
	where $d_{\text{eNb},13}=d_{\text{eNb},1} - d_{\text{eNb},3}$, $\delta_{13}= c (\Delta_{1}^\prime - \Delta_{3})$, and $\varepsilon_{13}=c(\varepsilon_{\text{SN},1} - \varepsilon_{\text{SN},3})$. Besides, $\Delta_{1}^\prime$ and $\Delta_{3}$ denote the measured time difference at  \( \mathbf{s}_1 \) and \( \mathbf{s}_3 \), respectively. Here, the errors \( \varepsilon_{12} \) and \( \varepsilon_{13} \) are assumed to be small and can be modeled as random noise with zero mean. 
	
	Besides, we can express the distance differences \( \Delta d_{1k} \) $k\in\{2,3\}$ in terms of the UE's position:
	\begin{align}
		\Delta d_{1k} &= \sqrt{(x_k - x_u)^2 + (y_k - y_u)^2} \nonumber\\
		&\quad-\sqrt{(x_1 - x_u)^2 + (y_1 - y_u)^2} \nonumber\\
		\Rightarrow\Delta d_{1k}&+\sqrt{(x_1 - x_u)^2 + (y_1 - y_u)^2} \nonumber\\
		&=\sqrt{(x_k - x_u)^2 + (y_k - y_u)^2} ,	\label{25} 
	\end{align} 
	Then, introducing $d_{\text{UE},1}=\sqrt{(x_1 - x_u)^2 + (y_1 - y_u)^2}$    and squaring both sides of \eqref{25}, we obtain the following set of hyperbola equation   
	\begin{align}
		&(x_u-x_1)(x_k-x_1)+(y_u - y_1)(y_k - y_1)+\Delta d_{1k}\times d_{\text{UE},1}\nonumber\\
		&=\frac{1}{2}\bigg[(x_k-x_1)^2+(y_k-y_1)^2-\Delta d_{1k}^2\bigg], \quad\quad k =2,3 \label{26}
	\end{align}  
	which represent hyperbolas on which the UE must lie.
	Writing \eqref{26} in matrix form gives
	\begin{align}
		{\bf G}{\bm \vartheta}={\bf h},
	\end{align} 
	where 
	\begin{align}
		\mathbf{G} &=
		\begin{bmatrix}
			x_2 - x_1 & y_2 - y_1 & \Delta d_{12} \\ 
			x_3 - x_1 & y_3 - y_1 & \Delta d_{13}
		\end{bmatrix},\\
		\mathbf{h} &= \frac{1}{2}
		\begin{bmatrix}
			(x_2^2+y_2^2) -(x_1^2+y_1^2) - \Delta d_{12}^2 \\ 
			(x_3^2+y_3^2)  -(x_1^2+y_1^2) -  \Delta d_{13}^2
		\end{bmatrix},
	\end{align} 
	and the parameter ${\bm \vartheta}=[x_u, y_u, d_{\text{UE},1}]^\text{T}$. Accordingly, we can formulate a LS problem \cite{KCho} to estimate the UE's position \( \mathbf{u} \). 
	\subsubsection{Formulating the LS Problem}
	With the system in the form ${\bf G}{\bm \vartheta}={\bf h}$, we can formulate the LS
	 problem to estimate ${\bm \vartheta}$, which is given as
	\begin{align} 
		\min_{\bm{\vartheta}} \quad J(\bm{\vartheta}) = \left| \mathbf{G} \bm{\vartheta} - \mathbf{h} \right|^2. \label{ls_problem} 
	\end{align} 
	\subsubsection{Solving the LS Problem}
	In the presence of measurement errors, we aim to find the parameter vector  ${\bm \vartheta}$ that minimizes the cost function \eqref{ls_problem}, which is a standard linear LS problem, which can be solved analytically. The optimal solution $\hat{\bm \vartheta}$ that minimizes $J(\bm{\vartheta})$ is obtained by solving the normal equations \cite{KCho}:
	\begin{align}
		\label{normal_equations} 
		\mathbf{G}^\mathrm{T} \mathbf{G} \hat{\bm{\vartheta}} = \mathbf{G}^\mathrm{T} \mathbf{h}.
	\end{align}
	Assuming $\mathbf{G}^\mathrm{T} \mathbf{G}$ is invertible, the solution is given by 
	\begin{align}
		\label{ls_solution} 
		\hat{\bm{\vartheta}} = \left( \mathbf{G}^\mathrm{T} \mathbf{G} \right)^{-1} \mathbf{G}^\mathrm{T} \mathbf{h}. \end{align}
	\section{Experimental Results}
\begin{figure}[t]
		\centering
		\includegraphics[width=0.8\linewidth]{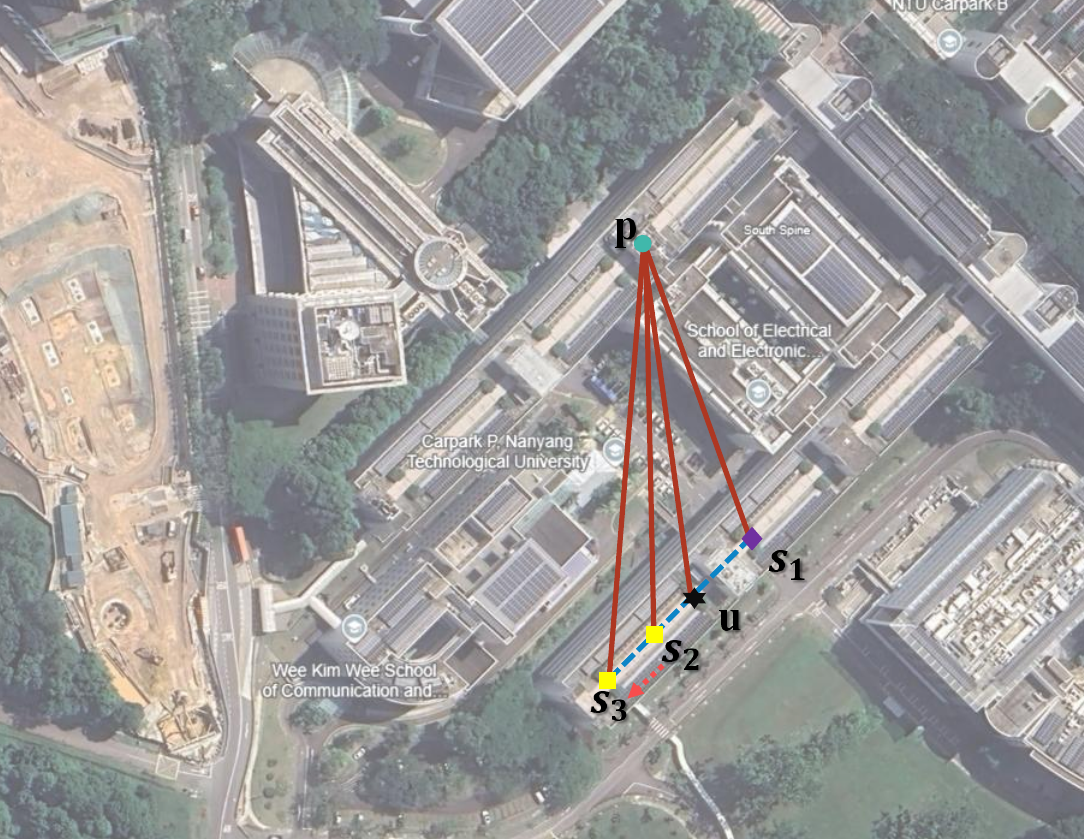}
		\caption{Satellite image of two sniffers experiment setup.}\label{ES}
	\end{figure}

	Our experimental setup consists of two X310 software-defined radios (SDRs) running LTE Sniffer v2.1.0 and an iPhone SE serving as the UE. As shown in Fig.~\ref{ES}, the eNb is located at the S1 EEE building rooftop of NTU, operated by M1 service providers. The whole setup is located within $\text{TA}=1$, where the internal boundary starts at 78.12\,m and ends at 156.24\,m, covering a total distance of 78.12\,m. The UE, SN 1, and SN 2 are positioned on the S2 EEE rooftop at distances of 114.70\,m, 109.70\,m, and 139.50\,m from the eNb, respectively. We positioned the UE in line of sight with the   eNb to ensure optimal signal reception. For each distance measurement of the UE, we reconnected six times to measure the distance over multiple connections, enhancing the accuracy of our measurements. 
	
	From Fig.~\ref{ES}, we can observe that to achieve the TDoA effect, we fix SN 1 at position \( \mathbf{s}_1 \) and move SN 2 from its initial position at \( \mathbf{s}_2 \) (139.50\,m from the eNb) to a new position at \( \mathbf{s}_3 \) (154.0\,m from the eNb).  
	\\\textbf{ {Remark 1}}: 
	\textit{Typically, implementing TDoA requires three sniffers; however, by moving one sniffer to another location and ensuring that during the same time period the two sniffers capture transmissions within the same subframe, we can achieve TDoA with only two sniffers.} 
	
\begin{figure}[t]
		\centering
		\includegraphics[width=1.0\linewidth]{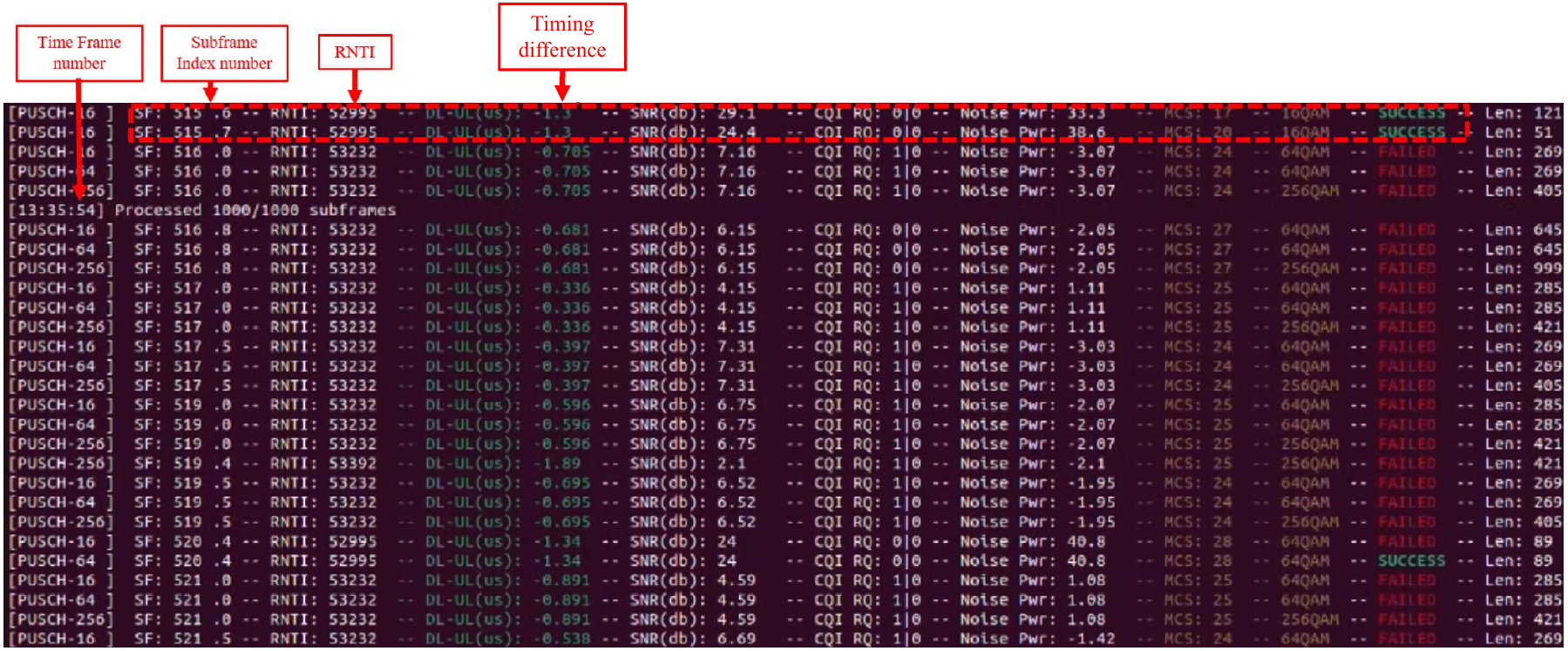}
		\caption{An example of terminal output from the sniffer v2.1.0.}\label{SN}
	\end{figure}

	Fig.~\ref{SN} shows the terminal output of the sniffer's logging interface. The sniffer first tunes to the same Physical Cell ID (PCI) and eNb as the UE. The output consists of subframe identification numbers, followed by the RNTI of each detected user in the area. Once the target UE is identified, the recording switches to logging UE-specific data. As shown in Fig.~\ref{SN}, each log entry includes the downlink-uplink (DL-UL) timing difference in microseconds, corresponding to the subframe timing difference for the individual UE. The signal-to-noise ratio (SNR) and channel quality indicator (CQI) are also printed for each UE, along with noise power. 
	
	The average number of active users during data recording under the same eNb ranges from 80 to 190. To capture the target radio network temporary identifiers (RNTIs) among multiple RNTIs, we placed our UE in radio resource control  (RRC) connection mode by toggling the network off and on to identify the target RNTI. Once identified, we initiated data recording for the UE. For extended data collection, we made a call to the UE to maintain the RNTI. Each dataset was recorded for 20 seconds, which is the minimum duration the UE retains the RNTI without an active call. In our experiment, UE position remained the same throughout the data recording. The plotted distance is calculated from the data points corresponding to the multiple delta values received during data collection. 
	
	We recorded data from SN 1 and SN 2 simultaneously to minimize the impact of random offset. To extract and match the delta value between the two sniffers, we first sorted the data with respect to time frames. Once the data was sorted by time frames, the next step was to align the same time frames for both sniffers to extract the delta value. Even within the same time frame, multiple delta values were present in the output. To identify the correct delta value for the same time frame, we extracted the subframe index number and matched it across both sniffer recordings, as illustrated in Fig.~\ref{SN}.
	\begin{figure}[t]
		\centering
		\includegraphics[width=0.85\linewidth]{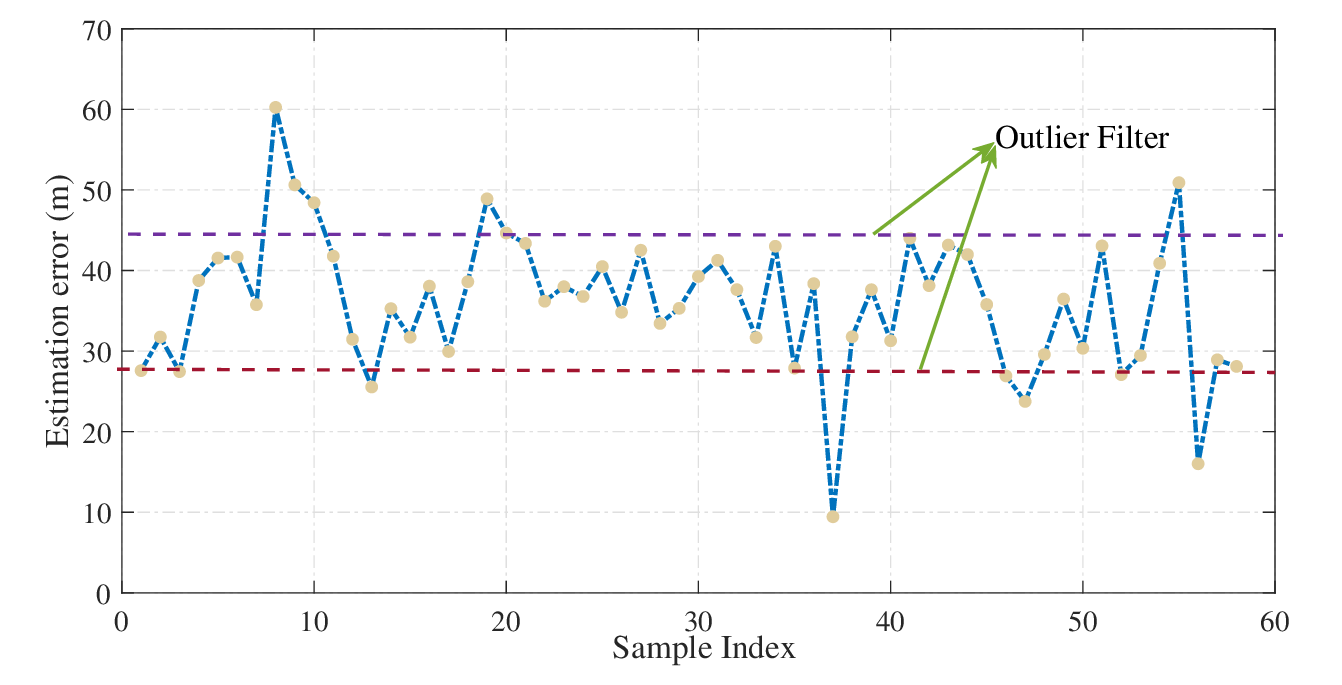}
		\caption{ToA measurement in 20dB.}\label{ToA_20dB}
	\end{figure}
	\begin{figure}[t]
		\centering
		\includegraphics[width=0.9\linewidth]{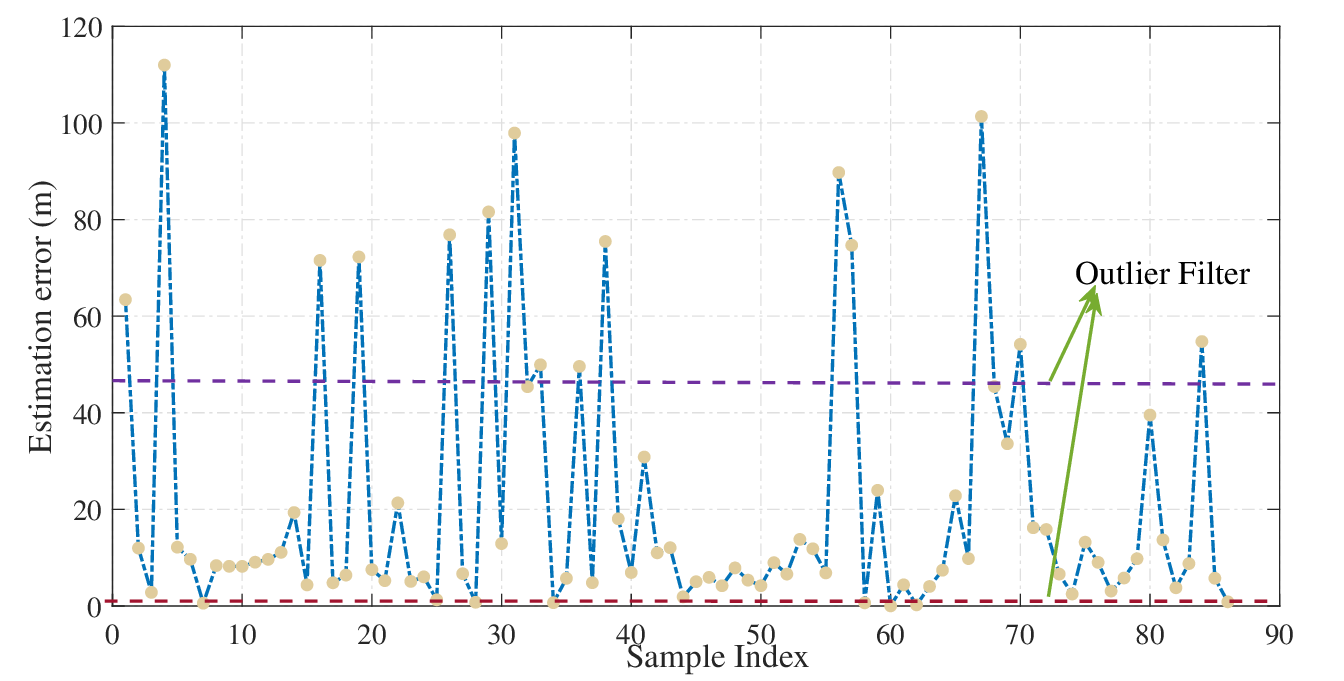}
		\caption{TDoA measurement in 20dB.}\label{TDoA_20dB}
	\end{figure}
	Fig.~\ref{ToA_20dB} shows the distribution of distance errors when using the ToA-based algorithm under 20 dB signal conditions. As observed from the figure, the estimation error can be as low as 10 m or less. To further enhance measurement accuracy, we applied an outlier filter using the 1-sigma rule as a threshold. Before applying the filtering, the estimated mean error is approximately 40 m. After filtering, the mean error is reduced to around 35 m, which aligns with the accuracy requirements of current 4G LTE localization standards. 
	\\\textbf{Remark 2:}  
	\textit{The 1-sigma outlier filter  identifies and removes data points that fall outside one standard deviation from the mean of the dataset. This technique is commonly used to improve data quality by filtering out extreme outliers, ensuring that only reliable measurements are considered for final estimation.}

	Although Fig.~\ref{ToA_20dB} demonstrates that the ToA-based scheme meets some localization requirements, it neglects hardware timing errors and measurement noise, resulting in limited accuracy. To improve the precision, we applied a TDoA-based  LS scheme. Fig.~\ref{TDoA_20dB} shows the distribution of distance errors using the TDoA-based LS scheme under 20 dB signal conditions. As illustrated in the Fig.~\ref{TDoA_20dB}, the application of the TDoA scheme reduces the estimated distance error to as low as $0.1$ m to $0.5$ m, achieving highly accurate localization. This demonstrates that, even with cost-efficient passive localization, we can achieve high precision within 4G networks, making it a practical solution for various applications. Additionally, the mean distance error before filtering was $21.44$ m, which was further reduced to $10.12$ m after filtering. This improved accuracy highlights the scheme's effectiveness, especially for applications in security-sensitive fields where precise localization is critical.
	
	\begin{figure}[t]
		\centering
		\includegraphics[width=0.9\linewidth]{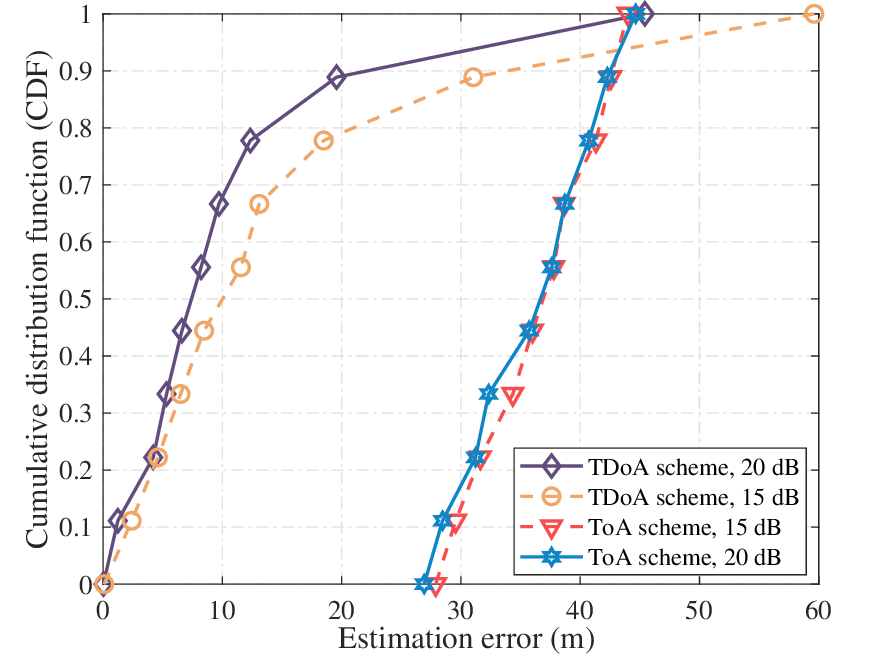}
		\caption{CDF of ToA and TDoA schemes at 15 dB and 20 dB.}\label{CDF}
	\end{figure}
	Fig. ~\ref{CDF} presents the cumulative distribution function (CDF) of the ToA and TDoA schemes under signal conditions of $15$ dB and $20$ dB. From Fig. \ref{CDF}, it is evident that increasing the SNR has little effect on improving the localization accuracy when using the ToA scheme, the results remain nearly unchanged regardless of the increase in SNR. In contrast, the TDoA scheme shows a significant improvement in accuracy as the SNR increases. Specifically, the estimation errors at the $80\%$ probability level are approximately $11$ m and $20$ me  for $20$ dB and $15$ dB, respectively, when using the TDoA scheme. 
	
	The observation from Fig. \ref{CDF} highlights the fact that the TDoA scheme benefits greatly from higher SNR, resulting in more accurate localization. \textit{Therefore, to achieve optimal localization accuracy, it is preferable to operate under high SNR conditions when applying the TDoA scheme.} However, in practical applications, it is also important to consider the trade-off between localization accuracy and detection cost. Higher accuracy generally requires the detection of more subframes, which may increase the overall computational and energy cost of the system. \textit{In most security-related applications, a signal condition of $15$ dB is sufficient to meet localization accuracy requirements, offering a reasonable balance between performance and operational efficiency without significantly increasing resource usage.} Thus, while higher SNR is ideal, the $15$ dB condition provides an acceptable compromise for most practical uses.
	
	\begin{table}[h!]
		\centering  
		\begin{tabular}{|c|c|c|c|}
			\hline
			\textbf{Scheme} & \textbf{Mean } & \textbf{RMSE } & \textbf{STD } \\ \hline
			ToA   & 36.83 m & 36.52 m & 4.79 m \\ \hline
			TDoA  & 13.99 m & 10.12 m & 9.65 m \\ \hline
		\end{tabular}
		\caption{Comparison for ToA and TDoA schemes.}\label{tab}
	\end{table}
	Table ~\ref{tab} provides a comparison of the mean, root mean square error (RMSE), and standard deviation (STD) for the ToA and TDoA schemes. It is evident that the TDoA scheme outperforms the ToA scheme in terms of localization accuracy. Specifically, the Mean and RMSE for TDoA are significantly lower, at $13.99$ mand $10.12$ meters, respectively, compared to $36.83$ m and $36.52$ m  for ToA. This demonstrates a notable improvement in estimation precision when employing the TDoA approach. Furthermore, the TDoA scheme exhibits a larger standard deviation ($9.65$ m) compared to ToA ($4.79$ m), indicating greater variability in the distance errors, which can be attributed to the additional measurements and processing required for the TDoA calculations. Overall, the TDoA scheme achieves better accuracy and lower RMSE, making it more suitable for applications that demand precise localization.
	\section{Conclusion} 
	We investigated a dual-sniffer passive localization system for  locating UE in densely populated urban environments. Two localization schemes were developed: a ToA based scheme and a TDoA based scheme.   Simulation results validated the effectiveness of the proposed TDoA-based scheme, demonstrating improved accuracy in UE positioning.
	
	\bibliographystyle{IEEEtran}
	\bibliography{myre}
\end{document}